# Cellular organization by self-organization: mechanisms and models for Min protein dynamics

Martin Howard[1] and Karsten Kruse[2]

[1]Department of Mathematics, Imperial College London, South Kensington Campus, London SW7 2AZ, UK.

[2]Max-Planck-Institut für Physik komplexer Systeme, Nöthnitzer Str. 38, 01187 Dresden, Germany.

We use the oscillating Min proteins of *Escherichia coli* as a prototype system to illustrate the current state and potential of modeling protein dynamics in space and time. We demonstrate how a theoretical approach has led to striking new insights into the mechanisms of self-organization in bacterial cells and indicate how these ideas may be applicable to more complex structure formation in eukaryotic cells.

**Introduction**

Self-organization has emerged in recent years as a vital concept in our understanding of subcellular architecture. As simple examples from physics and chemistry have shown, self-organizing behavior can spontaneously appear in systems with many interacting components, even when these components do not individually present the observed behavior (Cross and Hohenberg, 1993). Self-organization has a number of critical advantages for cellular systems, both in its flexibility and robustness (Misteli, 2001). In particular, self-organized structures can be both rapidly assembled and disassembled by simple modifications of the subcomponents, without the need for a dedicated machinery. At the same time, fluctuations in many of the components have only a limited effect on the structure as a whole. As we will argue in this minireview, the key to understanding self-organization is through the concept of a dynamic instability, where the intrinsic properties of a functional unit drive the system away from a disorganized state and towards a state of global spatiotemporal organization. For this to be achieved energy must, of course, be consumed and hence self-organization is an inherently non-equilibrium phenomenon. This should be contrasted with self-assembly processes (such as in viral capsids), which do not require energy input (Misteli, 2001).

When analyzing self-organizing systems, qualitative arguments and vague pictures of possible mechanisms often lack sufficient precision to decide whether a hypothesized mechanism is realized, or even possible at all. Instead, mathematical modeling and computational approaches provide the appropriate tools. For example, reaction-diffusion theory provides a well-developed conceptual framework for understanding pattern formation and morphogenesis in developing organisms (Koch and Meinhardt, 1994). Remarkably, as modeling has shown, self-organization often depends on only a small number of essential properties of the components. As a consequence, understanding a particular system may help to understand other systems that share essential features but which do not contain the same components. For example, in eukaryotic cell biology, understanding self-organization in microtubule-kinesin systems has important implications for the self-organization of actin-myosin systems. However, in order to be successful, theoretical approaches must be firmly rooted in experiments. As a first step, it might be sufficient to demonstrate that a certain mechanism is, in principle, possible, but eventually quantitative predictions must be obtained that allow the model to be tested against experiments.

In this minireview, we will illustrate the usefulness of a theoretical approach by focusing on the Min system in *Escherichia coli* as an example of a self-organizing system. We will discuss recent theoretical advances in understanding the self-organized subcellular distribution of the Min proteins in space and time. We will also emphasize how this work in bacterial biophysics has wide implications for self-organized structure formation in many cells, both eukaryotic and prokaryotic.

**From mechanisms to models: the Min system as a prototype**

The Min system in *E. coli* is a principle means for specifying the midcell position of the division plane, thus ensuring division into two equal sized daughter cells (Lutkenhaus, 2002). The first protein to assemble at the future division site is the tubulin homologue FtsZ, which polymerizes into a ring shaped structure at midcell (Lutkenhaus, 2002). The position of the FtsZ ring is negatively regulated by the distribution of DNA, leaving three potential sites of division: at midcell and close to each of the cell poles. At the poles, division is suppressed by the action of the Min proteins: MinC, MinD, and MinE. MinD both binds to the membrane and is able to form a complex with MinC, which it recruits to the membrane where it can act efficiently to inhibit assembly of the FtsZ ring. MinE is also recruited by MinD to the membrane but once there, acts to expel membrane bound MinCD. Importantly, MinD is an ATPase: in its ATP form it binds to the membrane from where its unbinding is stimulated by MinE and requires energy from ATP hydrolysis (Hu et al., 2002). Hence, the dynamics of the Min system is inherently out of equilibrium, allowing the possibility of self-organizing behavior. Indeed, fluorescent imagery experiments have uncovered remarkable spatiotemporal oscillations of all the Min proteins, with a period of a minute or less (Hale et al., 2001; Hu and Lutkenhaus, 1999; Raskin and de Boer, 1999). During the course of the oscillations, the proteins alternate between the two cell halves. As a result, MinC inhibits the formation of the FtsZ ring close to the cell poles and cell division is thus directed to the cell midplane. Interestingly, while the oscillations are critical for the function of MinC, it apparently does not contribute to the oscillation mechanism, since the MinD/MinE oscillations persist even without MinC. A schematic representation of the MinD/MinE oscillations is presented in Fig. 1.

Until recently, the origin of the Min oscillations has remained mysterious: what is the cause and driving mechanism of the dynamics? This puzzle has turned out to be hard to resolve, especially as the interlinked interactions of the Min proteins make it difficult to reliably predict their behavior from qualitative pictures alone. However, a solution has recently been obtained using a mathematical approach (Howard et al., 2001; Huang et al., 2003; Kruse, 2002; Meinhardt and de Boer, 2001). The Min system is ideally suited for such an analysis as it displays complex spatiotemporal behavior while consisting of only a few components and interactions, most of which are well characterized by a decade or more of intensive experimental study. In essence, the mathematical models explain the Min oscillations using only the interactions between MinD, MinE, and the cytoplasmic membrane: MinD binds to the membrane, recruits MinE, which then induces detachment of MinD and itself from the membrane. Beyond these core assumptions, however, the models differ: continuous protein synthesis and degradation is crucial for the model presented in (Meinhardt and de Boer, 2001), while in (Howard et al., 2001) membrane-bound MinE is assumed to reduce the attachment rate of MinD. For the models in (Huang et al., 2003) and (Kruse, 2002) the formation of MinD aggregates, by a one-step or two-step process, respectively, is essential.

In spite of these detailed differences, the fundamental mechanism behind the oscillations is the same in all cases, namely a dynamic instability of a homogenous distribution of the Min proteins. This instability can be analyzed mathematically and follows directly from diffusion plus the few core properties of the Min proteins sketched above. In all cases computer simulations of the models show oscillatory Min dynamics with a time-averaged midcell minimum of the MinCD concentration in good agreement with experiment. As we have already mentioned, a dynamic instability is the hallmark of self-organization, where spatiotemporal patterns emerge due to the inherent properties of a functional unit. Note that it is important to distinguish here between a fundamental mechanism (in this case a dynamic instability) and its subsequent implementation into a specific model, of which there will likely be several possible variations compatible with the available experimental data at a given time. For the Min system, the existing models have firmly established the fundamental mechanism of a dynamic instability, leading to spontaneous Min protein oscillations. The need now is to identify key experiments that will be able to more carefully discriminate between the models, testing both the fundamental assumptions of each model and also probing model predictions. For example, the models vary in their detailed predictions for the oscillation periods as a function of the protein concentrations and bacterial length. More quantitative experiments on this issue would provide a powerful source of comparison. The Min proteins are also now known to polymerize into helical filaments on the cell membrane (Hu et al., 2002; Shih et al., 2003), an important feature not yet captured by the models. Incorporating this effect is an obvious next step for modeling in a continued dialogue between experiment and theory. The final goal of this partnership would be a fully quantitative model of the Min protein dynamics, which could then serve a prototype example for the analysis of other self-organizing modules in the cell.

**Applications far and wide**

An increasing number of important prokaryotic proteins are now known to undergo precisely targeted, dynamic localization. Hence the remarkable dynamics of the Min system are by no means unique. Clearly the old idea that bacteria are simply featureless bags of enzymes must be consigned to history (Errington, 2003; Lutkenhaus, 2002; Shapiro et al., 2002). Mathematical modeling of these systems is at an early stage compared to the Min dynamics, but is likely to be similarly revealing. The ParA protein involved in segregation of the virulence plasmid pB171 in *E. coli* oscillates coherently in nucleoid regions (Ebersbach and Gerdes, 2001; Hunding et al., 2003). In *Bacillus subtilis* the Soj protein implicated in chromosome organization and as part of a sporulation checkpoint oscillates from nucleoid to nucleoid in the presence of the Spo0J protein (Marston and Errington, 1999; Quisel et al., 1999). In *Caulobacter crescentus* the signaling protein DivK has recently been shown to shuttle from pole to pole undergoing phosphorylation at one end and then dephosphorylation at the other (Matroule et al., 2004). Following completion of cytokinesis, diffusion is no longer possible between the old cell poles, meaning that DivK is then entirely converted to its phosphorylated form in one daughter cell, and its dephosphorylated form in the other. This difference is then used to drive different development programs between the two daughter cells. Furthermore, even apparently static objects such as the FtsZ ring, have now been shown to be highly dynamic structures, undergoing remodeling on a time scale of seconds (Anderson et al., 2004). Finally, the cell division location in *B. subtilis* is controlled by a rather different system from *E. coli*. *B. subtilis* lacks the MinE protein and instead possesses an unrelated protein called DivIVA which serves to anchor the MinCD complex to the cell poles. This localization, which might also involve high turnover, likely involves polar geometric cues (Howard, 2004), and is a reminder of how bacteria can solve the same problem (accurate cell division) in different ways, while still retaining the recurring theme of dynamic protein localization. In all likelihood numerous fascinating examples of dynamic proteins in bacteria remain to be discovered.

The systems discussed so far all concern the localization of proteins in bacteria. Is there anything to be learned from these systems for eukaryotic cells? To answer this question, we note that MinD and FtsZ are components of the bacterial cytoskeleton. FtsZ is an analog of tubulin (Lowe and Amos, 1998), while, in eukaryotes, MinD plays an important part in the fission of plant plastids, similar to its role in bacteria. In addition, other cytoskeletal proteins have been discovered in bacteria: for example MreB and Mbl are analogs of actin and play a role in maintaining cell shape and polarity (Gitai et al., 2004; van den Ent et al., 2001), while crescentin in *C. crescentus* is the first prokaryotic analog of intermediate filaments (Ausmees et al., 2003). These proteins form structures similar to those found in eukaryotes. Notably the FtsZ ring resembles the contractile ring formed by actin filaments in the late stages of cell division in eukaryotic cells, and ParM, another actin analog, forms a structure reminiscent of the mitotic spindle to separate the low copy number plasmid R1 (Moller-Jensen et al., 2003).

Theoretical analysis of the dynamics of the eukaryotic cytoskeleton has so far focused on the interaction of filaments and motors. Initial studies, motivated by in vitro experiments on the self-organization of microtubules and kinesin, have simulated microscopic models for motors and filaments (Nedelec, 2002). They have reproduced the self-organized structures observed in vitro and yielded valuable insights into possible building blocks of stable spindle-like structures and into the mechanisms for their formation via dynamic instabilities. Phenomenological descriptions, similar in the spirit to the Min models described above, have revealed possible dynamic behaviors that rely solely on the polar nature of the filaments as well as on the directional motion of motors (Kruse et al., 2004; Kruse and Jülicher, 2000; Kruse and Jülicher, 2003; Liverpool and Marchetti, 2003). In particular, the existence of traveling waves has been predicted (Kruse and Jülicher, 2003). Such waves have recently been observed in *Dictyostelium discoideum* (Bretschneider et al., 2004). Spontaneous oscillations of many interacting molecular motors have opened up new explanations for the beating of eukaryotic cilia and flagella (Brokaw, 1975; Camalet and Jülicher, 2000). These explanations fundamentally rely on oscillatory dynamic instabilities. Applying these theories to cilia in the hair bundle of auditory hair cells, and exploiting the properties of a system close to an oscillatory instability, has convincingly linked several formerly unconnected characteristics of vertebrate hearing (Camalet et al., 2000; Eguiluz et al., 2000).

From these works some general principles underlying the organization of the eukaryotic cytoskeleton start to emerge. Notably, dynamic instabilities seem to play an important role and might be used to switch between different cellular behaviors. As discussed in detail for the Min system, exactly the same principles apply to bacteria. As far as we know today, though, there is one crucial difference between the prokaryotic and eukaryotic cytoskeleton, as - with the exception of RNA-polymerases – there are no linear motors in bacteria. Transport of molecules therefore seems to occur exclusively by diffusion, which in view of the small size of bacteria is very effective. In fact, a molecule typically needs only around a second to sample the whole cell. In eukaryotic cells, which are usually much larger, molecules are commonly transported by motors following the tracks provided by microtubules and actin filaments. It will be interesting to see in the future how both eukaryotic and prokaryotic cells use strategies involving dynamic instabilities, but based on different transport mechanisms, to solve common problems.

**Conclusions**

Theoretical analysis has already led to some profound insights into the mechanisms for self-organization and dynamic protein localization in cells. The models used in this context are inevitably more sophisticated than those employed in some other areas of biological modeling as spatial variation must be included. Nevertheless current models, for example for the Min system, are still quite crude and further effort is needed to increase their level of realism so that a full synthesis can be achieved with experiments. A further element that needs urgent attention is stochastic effects, which are now widely recognized to be important in cell biology (Paulsson, 2004). However, the investigation of stochastic effects in spatially extended systems is just beginning. One study does already exist of fluctuations in the Min system where it was shown that stochasticity does not compromise the accuracy of cell division, provided the copy numbers of the Min proteins are not too low (Howard and Rutenberg, 2003). Nevertheless, we believe that much more remains to be investigated in this direction.

We would like to conclude by making some remarks of a more general nature about mathematical modeling in biology. In particular we emphasize that successful modeling often requires the marriage of a "top down" concept, such as self-organization, to provide a theoretical framework, with a "bottom up" approach to ensure that the analysis accords with experiment, and can also make firm predictions. Using either approach in isolation is dangerous, as one either gets swamped in details or gets a mechanism to work that may not be relevant to the biological system at hand. In order to narrow things down and to start modeling on the right track one clearly needs as much data as possible, which in the context of cellular self-organization must include spatiotemporal data (e.g. dynamic fluorescence localization patterns). Even with this benefit, it is often difficult to generate a unique model, although, at least initially, one may be satisfied with a proof of principle for the overall mechanism rather than finding a single all-encompassing model. Later on, further experiment/theory iterations can be used to generate more realistic models with better experimental agreement but with the same underlying mechanism. If used appropriately modeling can then form a powerful complementary tool to experiment.

In essence, the power of modeling lies in boiling a system down to its bare essentials and thus exposing the minimal ingredients, which will explain the observed phenomena in a way, which is often hard to identify experimentally. As we have seen, the union of even quite simple interactions can yield surprisingly complex behavior, where the whole is definitely more than the sum of the parts. In these cases theory will be vital for an understanding of emergent properties that cannot be reliably identified from qualitative pictures: self-organization really does demand a mathematical description.

Matroule, J.Y., H. Lam, D.T. Burnette, and C. Jacobs-Wagner. 2004. Cytokinesis monitoring during development; rapid pole-to-pole shuttling of a signaling protein by localized kinase and phosphatase in Caulobacter. *Cell*. 118:579-90.

Meinhardt, H., and P.A. de Boer. 2001. Pattern formation in Escherichia coli: a model for the pole-to-pole oscillations of Min proteins and the localization of the division site. *Proc Natl Acad Sci U S A*. 98:14202-7.

Misteli, T. 2001. The concept of self-organization in cellular architecture. *J Cell Biol*. 155:181-5.

Moller-Jensen, J., J. Borch, M. Dam, R.B. Jensen, P. Roepstorff, and K. Gerdes. 2003. Bacterial mitosis: ParM of plasmid R1 moves plasmid DNA by an actin-like insertional polymerization mechanism. *Mol Cell*. 12:1477-87.

Nedelec, F. 2002. Computer simulations reveal motor properties generating stable antiparallel microtubule interactions. *J Cell Biol*. 158:1005-15.

Paulsson, J. 2004. Summing up the noise in gene networks. *Nature*. 427:415-8.

Quisel, J.D., D.C. Lin, and A.D. Grossman. 1999. Control of development by altered localization of a transcription factor in B. subtilis. *Mol Cell*. 4:665-72.

Raskin, D.M., and P.A. de Boer. 1999. Rapid pole-to-pole oscillation of a protein required for directing division to the middle of Escherichia coli. *Proc Natl Acad Sci U S A*. 96:4971-6.

Shapiro, L., H.H. McAdams, and R. Losick. 2002. Generating and exploiting polarity in bacteria. *Science*. 298:1942-6.

Shih, Y.L., T. Le, and L. Rothfield. 2003. Division site selection in Escherichia coli involves dynamic redistribution of Min proteins within coiled structures that extend between the
17

**Figure Legend**

**Figure 1.** Schematic representation of MinD/MinE oscillations in *E. coli*. Three successive time instants are illustrated. (**a**) MinD forms a membrane-bound helix in one half of the cell; MinE is associated with this structure, predominantly towards the center. (**b)** MinE stimulates detachment of MinDE from the membrane, thereby freeing the cell center for cell division. MinD/MinE diffuse in the cytosol, and, driven by the dynamic instability, **(c)**, MinD /MinE form a helix at the opposite end of the cell, and the process repeats. Note that the detailed structure of the MinD filaments as well as the exact location of MinE with respect to the filaments are currently unknown.

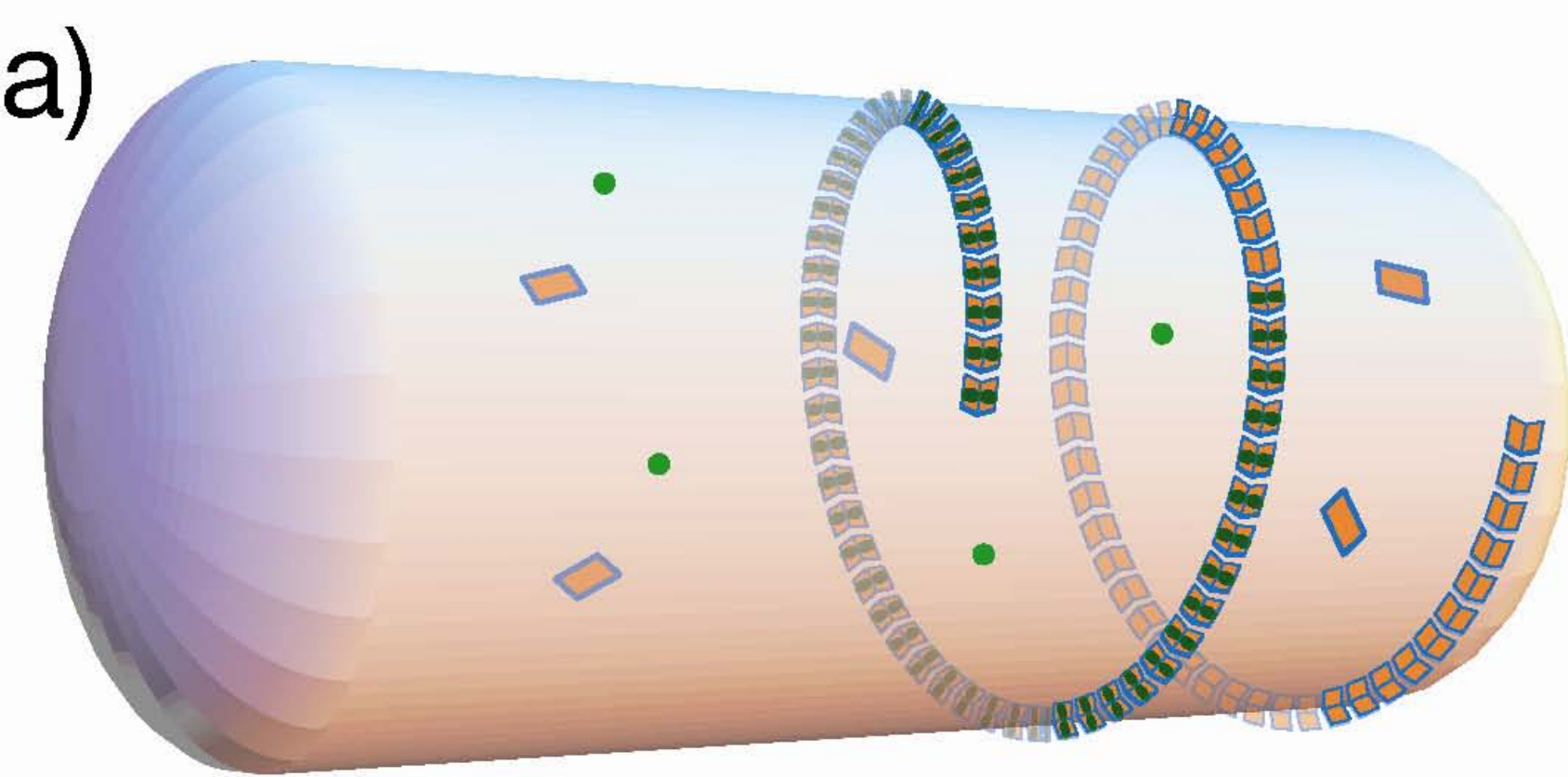

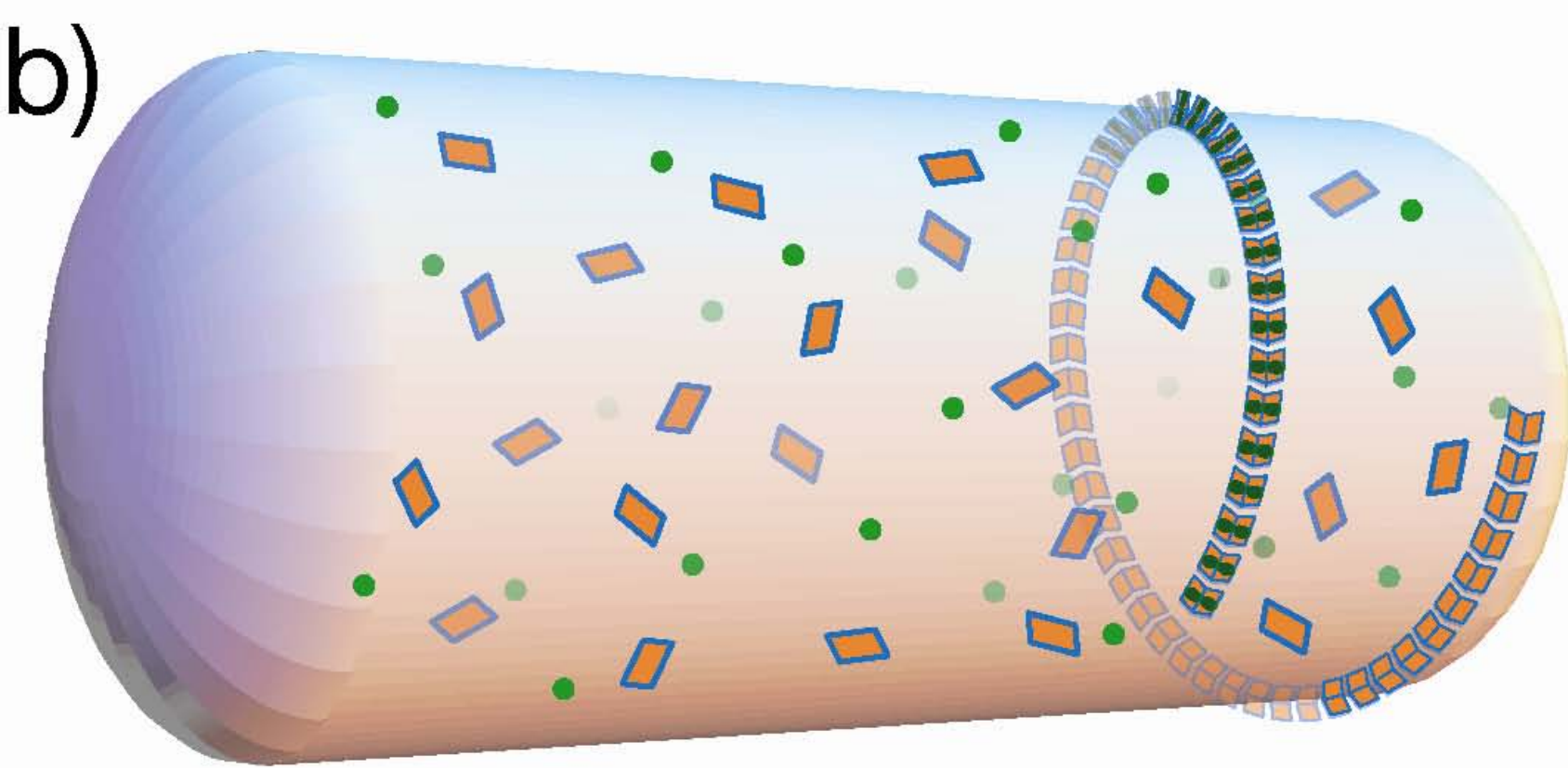

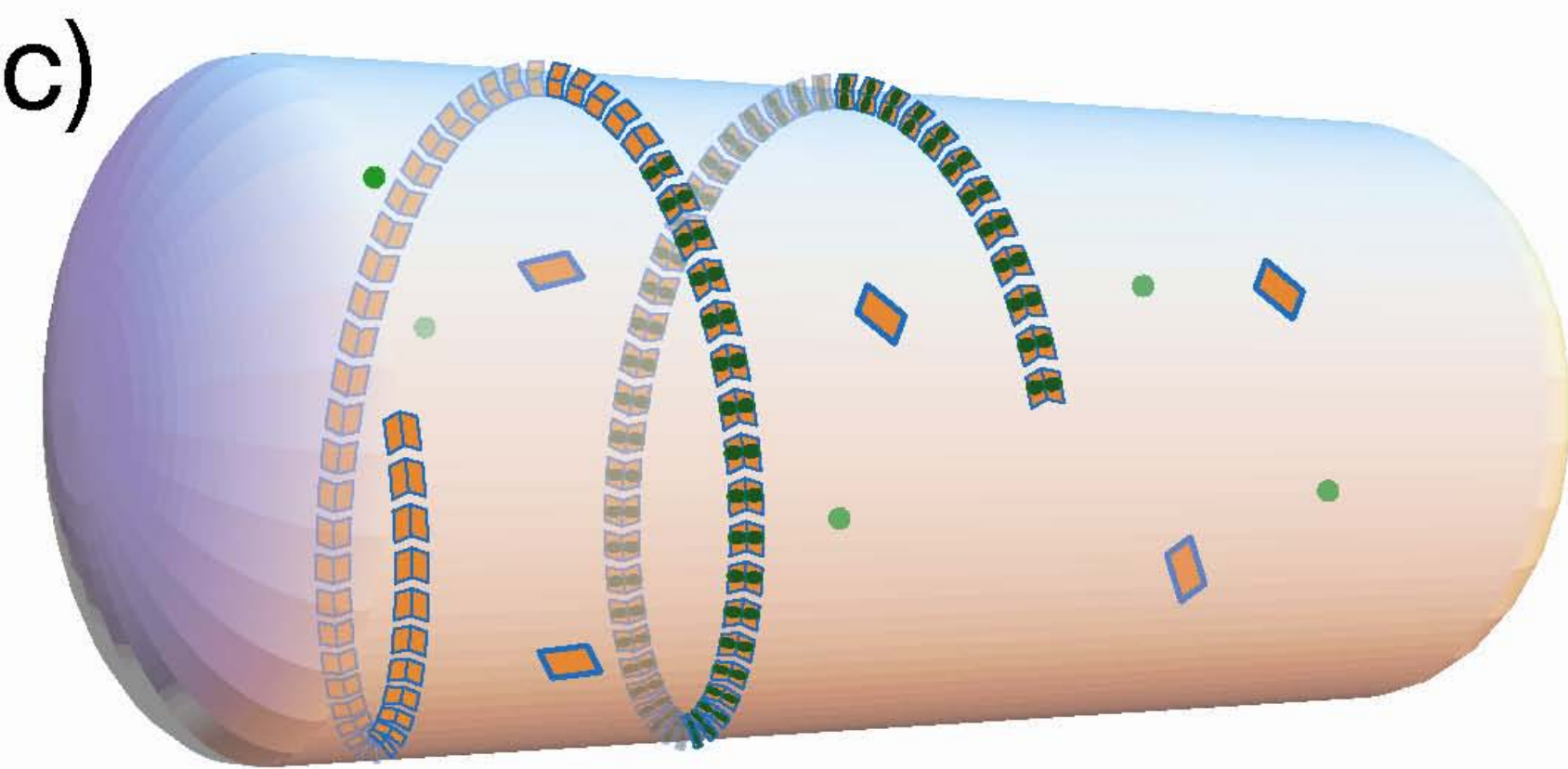

| MinD (orange parallelogram) | MinD-filament (chevron) |
| MinE (green dot) | MinDE-complex |